\def   \ni {\noindent}

\def   \ssk {\vskip  5truept}
\def   \sk  {\vskip 10truept}
\def   \bsk {\vskip 15truept}

\def   \newline {\hfil\break}

\documentclass{article}
\usepackage{epsfig}

\begin{document}

\hsize 5truein
\vsize 8truein
\font\abstract=cmr8
\font\keywords=cmr8
\font\caption=cmr8
\font\references=cmr8
\font\text=cmr10
\font\affiliation=cmssi10
\font\author=cmss10
\font\mc=cmss8
\font\title=cmssbx10 scaled\magstep2
\font\alcit=cmti7 scaled\magstephalf
\font\alcin=cmr6
\font\ita=cmti8
\font\mma=cmr8
\def\ref{\par\noindent\hangindent 15pt}
\null


\title{\ni NUCLEAR PHOTODISINTEGRATION AT\\ HIGH REDSHIFTS BY GAMMA-RAY
BLAZARS}

\bsk \bsk
\author{\ni M.~Cass\'e$^{1,2}$, E.~Vangioni--Flam$^{2}$ and R.~Lehoucq$^{1}$}
\bsk
\affiliation{\ni 1) CEA/DSM/DAPNIA/Service d'Astrophysique, France \\
                 2) Institut d'Astrophysique de Paris, CNRS, France
}
\bsk
\baselineskip = 12pt

\abstract{\ni ABSTRACT: We analyse the photodisintegration of nuclei 
near the center of blazars, in particular PKS~0528+134. We show that 
this process could explain the recent observation of a high Al/Si 
ratio in a cloud on the line of sight of a distant quasar and we 
calculate the effect of this gamma irradiation on the abundance of 
deuterium, an isotope of the highest cosmological interest since it is 
sensitive to the baryon density of the universe. The process has 
interesting consequences and deserves to be inserted in a consistent 
scenario of the early evolution of the universe.}

\sk
\baselineskip = 12pt

\bsk
\baselineskip = 12pt

\ni 1. INTRODUCTION
\ssk

\ni Blazars are extremely intense sources of gamma-rays and in their 
close vicinity, the abundance of certain nuclei could be distorted by 
photodisintegration. This process has been applied to AGN 
(e.g. Boyd, Ferland and Schramm 1989) but never explicitly to 
blazars. The aim of this work is to stress the interest of this 
mechanism and to make a link between gamma-ray astronomy, 
nucleosynthesis and cosmology following a line of reasoning close to 
that of Gnedin and Ostriker (1992, 1995), but more observationally 
oriented (von Montigny et al. 1995, Mc Naron--Brown et al. 1995). 
The Big Bang connexion is deuterium whose abundance could be modified 
by photodisintegration in high redshift clouds where it is observed by 
the most powerful instruments as the HST and the KECK telescope 
(Wampler et al. 1996, Webb et al. 1997, Tytler et al. 1996, 
Vidal--Madjar et al. 1998, Songaila 1997). Contrary to Sigl et al. 
(1995) we don't care of $^3$He since it is unobservable in cosmological 
clouds. Indeed the measurements of the abundances in high redshift 
absorbers deliver a wealth of data (Petitjean and Charlot 1998). 
Among them, one is particularly intriguing: two clouds on the line of 
sight of a quasar at $z = 1.94$ show very different Al/Si ratios 
(Ganguly et al. 1998). Cloud A has a normal pop II abundance for its 
age (Al/Si$\,\approx 10^{-2}$). However Cloud B show a strong Al 
overabundance with respect to solar (Al/Si$\,\approx 1$). No stellar 
source can explain this ratio. For instance the production ratio in 
core collapse supernova is about 0.1, close to the solar one. The high 
Al/Si ratio observed, reminiscent of the one observed in galactic 
cosmic rays is certainly of non-thermal origin. We surmize that it is 
due to the removal of one nucleon from $^{28}$Si by protons or other 
particles. Here we favor high energy photons since blazars are very 
copious sources of gamma-rays. The spallation case will be described 
elsewhere. Accordingly, the production mechanism should be 
$^{28}$Si($\gamma$,n)$^{27}$Si$\,\rightarrow\,^{27}$Al and 
$^{28}$Si($\gamma$,p)$^{27}$Al. The ($\gamma$,n) cross section are 
well measured. The ($\gamma$,p) one is only estimated. On the other 
hand, the D/H ratio measured in high $z$ clouds is very dispersed 
(ranging from $3\times 10^{-5}$ to $2\times 10^{-4}$, Vidal--Madjar et 
al. 1998), and we are inclined, provided this 
dispersion is real, to explore the idea that it is due to 
photodisintegration of He and D in the vicinity of blazars. Apart the 
cross sections of the various reactions of interest, 
the main parameters of the problem are the flux at the source above 
the energy threshold of the relevant reactions, the duration of the 
emission and the average spectrum of the photons.

\bsk
\ni 2. Al AND D PRODUCTION/DESTRUCTION

\ssk
\ni 2.1 Cross sections
\ssk

\ni As far as Al and D are concerned, the main production and 
destruction cross sections, with their associated energy threshold 
 are the following:

\begin{eqnarray}
\nonumber \gamma +  ^{4}\mathrm{He}&\rightarrow&\mathrm{D}\,(E >
20\,\mathrm{MeV}) \\
\nonumber \gamma + ^{28}\mathrm{Si}&\rightarrow&\mathrm{n} +
^{27}\mathrm{Si}\,\rightarrow\,^{27}\mathrm{Al} \\
\nonumber \gamma + ^{28}\mathrm{Si}&\rightarrow&\mathrm{p} +
^{27}\mathrm{Al}\,(E > 17-10\,\mathrm{MeV}) \\
\nonumber \gamma +      \mathrm{D} &\rightarrow&\mathrm{p} + \mathrm{n}\,(E
> 2\,\mathrm{MeV}) \\
\nonumber \gamma + ^{27}\mathrm{Al}&\rightarrow&\mathrm{X}\,(E > 10
\mathrm{MeV}) \\
\nonumber \gamma + ^{28}\mathrm{Si}&\rightarrow&\mathrm{X}\,(E > 17
\mathrm{MeV})
\end{eqnarray}

The reaction $^{4}$He($\gamma$,D)D is unsignificant.The production of 
D by $^{3}$He will be considered in future work. D/He is very sensitive 
to the photon flux between 2 and 30 MeV and Si/H to the flux between 
20 and 30 MeV.

\ssk
\ni 2.2 Spectra and fluxes
\ssk

\ni Recent observations by the three CGRO instruments, EGRET, OSSE and 
COMPTEL, have revealed a new class of extragalactic gamma ray sources 
associated to blazars AGN (von Montigny et al. 1995, Mc Naron--Brown 
et al. 1995, see Dermer, this meeting, for a review). Blazars are 
flat spectrum radio sources with jets closely aligned with our line of 
sight. In short and violent flaring phase the accelerated particles 
are injected in the form of plasma blobs along a preferential axis, 
forming a jet. We are not concerned by the detailed physics of these 
fascinating objects but by the long term (or statistical average) of 
their properties. Blazar spectra above 30 MeV,can be approximated by 
a power law ($E^{-1.4}$ to $E^{-2.7}$). EGRET, COMPTEL and OSSE 
observations combined show a flattening of the spectrum in the MeV 
range, above 1 to 50 MeV. The change in slope is between 0.5 and 1. 
We have shifted the observed spectra back to the sources, taking into 
account the cosmological effects through a model of the universe 
implying reasonable parameters (expansion rate, matter density and 
cosmological constant), which will be described in a forthcoming paper 
(Lehoucq et al. 1999).

With all these parameters at hand, we can calculate the effect of 
gamma-ray irradiation on a cloud located at a distance $R$ of the 
central engine of a blazar during a time $t$, then select a $(R,t)$ 
couple and finally estimate the corresponding alteration of D at a 
fixed Al/Si enhancement, here the one observed by Ganguly et al. 
(1998), say by a factor about of 100.

\bsk
\ni 3. RESULTS
\ssk

\ni As a reference model we have taken PKS~0528+134 associated to a 
cosmological model with $h_{100} = 0.7$, $\Omega_{m} = 0.3$ and 
$\Lambda = 0$. It is a very strong gamma ray source located at a 
redshift of $z = 2.06$, with a rather flat power law spectrum ($\Phi 
\propto E^{-1.5}$) which steepens above 17 MeV. Note that at the 
source, the break is shifted to about 60 MeV due to multiplication by 
the factor $(1 + z)$. As shown in figure 1, to get an Al/Si ratio of 
1 as observed in cloud B, starting from a ratio of $10^{-2}$ (cloud A) 
we need an irradiation time $t$ of $6.7\times 10^{3}$ year at a 
distance of 1 parsec. This result can be scaled to any other distance 
considering that t is inverselly proportional to the square of $R$. 
It is worth noting that short times scales, by astronomical standards, 
are involved. The sensitivity of the results to the parameters are 
shown in Tables~1 and 2, other spectra (for the same number of 
photons) ($E^{-1.4}$ and $E^{-2}$), two different primordial D/H 
ratios ($10^{-4}$ and $4\times 10^{-5}$) and a rather extreme 
cosmology ($\Omega_{m} = 0.1$ and $\Lambda = 0.9$) have been 
considered for comparison with the standard case. It appears that the 
most influencial parameter is the spectral index. Flat spectra are 
more efficient than steep ones to produce the required effect. Note 
that D can be both produced or destroyed depending on the shape of the 
spectrum, this by modest amounts (less than 4). Cosmological effects 
introduce at most a factor of about 2 at $z = 2$. We think that we 
have demonstrated the interest of photodisintegration for the 
understanding of the abundances of nuclei in high redshift clouds 
(Cass\'e and Vangioni--Flam 1998). This mechanism remains to be 
integrated in a consistent scenario of cosmic nuclear evolution of the 
kind developed by Gnedin and Ostriker (1992, 1995). The elaboration 
of this great perspective should start, most modestly, from the study 
of irradiation of falling matter on the central black hole feeding 
blazars, its ejection in relativistic blobs, and its subsequent 
condensation in the form small intergalactic clouds.

\begin{table}[tbp]
    \centering
    \caption{TABLE 1: Sensitivity of the results to the physical parameters.}
    \begin{tabular}{|l|c|c|c|}
        \hline
        Spectrum & Irradiation time (yr) & 
	D/$D_{p}$ ($D_{p} = 4\times10^{-5}$) & D/$D_{p}$ ($D_{p}$ = $10^{-4}$)  \\
        \hline
        PKS~0528+134 & $6.7\times 10^3$ & 4 & 1.6  \\
        $E^{-1.4}$ & $1.3\times 10^4$ & 1.3 & 0.5  \\
        $E^{-2}$ & $1.8\times 10^5$ & 2.5 & 0.5  \\
        \hline
    \end{tabular}
    \label{tbl:sensitivity}
\end{table}

\begin{table}[tbp]
    \centering
    \caption{TABLE 2: PKS~0528+134 case.}
    \begin{tabular}{|l|c|c|}
        \hline
        Cosmological models & Irradiation time (yr) & 
	D/$D_{p}$ ($D_{p} = 10^{-4}$)  \\
        \hline
        $\Omega_{m} = 0.3\,\,\Lambda = 0$ & $6.7\times 10^3$ & 1.6  \\
        $\Omega_{m} = 0.1\,\,\Lambda = 0.9$ & $3.4\times 10^3$ & 1.6  \\
        \hline
    \end{tabular}
    \label{tbl:PKS}
\end{table}

\begin{figure}
\centerline{\psfig{file=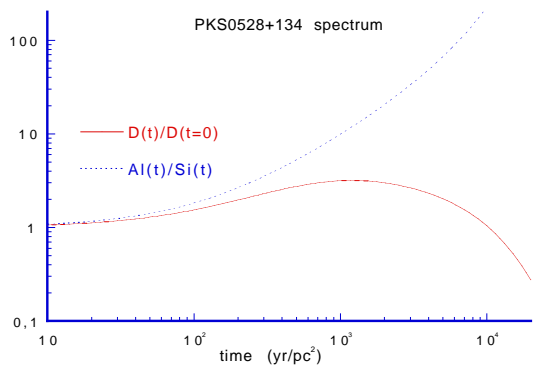, width=7cm}}
\caption{FIGURE 1. Evolution of Al/Si and D/H for the standard case
(PKS~0528+134 associated to a cosmological model with $h_{100} = 70$,
$\Omega_{m} = 0.3$ and $\Lambda = 0$.}
\end{figure}

\bsk
\ni 4. CONCLUSIONS
\ssk

\ni The recent observation of a large Al/Si ratio in a cloud on the 
line of sight of a distant quasar seem to justify the 
photodisintegration mechanism. The high Al/Si observed could be 
explained by gamma irradiation near a blazar core during 
$10^{4}/R_\mathrm{pc}^{2}$ years. The final Al/Si ratio depends 
weakly on the cosmology adopted (at least up to $z = 4$). A flat 
spectrum ($E^{-1.4}$) is favoured. The final abundance of D is 
sensitive to the source spectrum between 2 and 20 MeV and depends on 
the initial D/H ratio adopted. D can be circunstantially destroyed or 
produced by quite moderate factors however ($< 4$). These variations 
could explain the scattering of the observed D/H on line of sight of 
remote quasars, if real. As a conclusion, photodisintegration is 
worth integrating in a consistent cosmological/astrophysical scenario.

\bsk
\baselineskip = 12pt
{\abstract \ni ACKNOWLEDGMENTS: We thank warmly Pr. Laget from the Service de
Physique Nucl\'eaire (CEA/DSM/DAPNIA) in Saclay fo his help in the choice of
photodisintegration cross sections. This work has been done under PICS~319,
CNRS.}

\bsk
\baselineskip = 12pt

{\references \ni REFERENCES
\ssk

\ref Boyd, R.N., Ferland, G.J. and Schramm, D.N., 1989, ApJ 336, L1

\ref Carswell, R.F., Weymann, R.J., Cooke, A.J. and Webb, K.J., 1994, MNRAS
268, L1

\ref Cass\'e, M. and Vangioni--Flam, E., 1998, in "Structure and evolution of
the intergalactic medium from QSO absorption lines", ed. P. Petitjean and S.
Charlot, ed. Fronti\`eres, p. 331

\ref Ganguly, R., Churchill, W. and Charlton, J.C., 1998, ApJ 498, L103

\ref Gnedin, N.Y. and Ostriker, J.P., 1992, ApJ 400, 1

\ref Gnedin, N.Y., Ostriker, J.P., 1995, ApJ 438, 40

\ref Lehoucq, R. et al., 1999, in preparation

\ref MacCormick M. et al., 1996, Phys. Rev. C, 52, 41

\ref McNaron-Brown, K. et al., 1995, ApJ 451, 575

\ref Petijean P. and Charlot, S. 1998 "Structure and evolution of the
intergalactic medium from QSO absorption lines", ed. P. Petitjean and S.
Charlot, ed. Fronti\`eres

\ref Pywell, R.E. et al 1983, Phys.Rev. C 27, 960
\ref Songaila, A., 1997, in "Structure and evolution of the intergalactic
medium
from QSO absorption lines", ed. P. Petitjean and S. Charlot, ed.
Fronti\`eres, p. 339

\ref Sigl, G., Jedamzik K., Schramm D.N. and Berezinsky V.S. 1995, Phys. Rev.
 D, 52, 6682

\ref Tytler, D., Fan, X.M., and Burles, S., 1996, Nature 381, 207

\ref Veyssiere et al., 1974, Phys. Nuc. A227, 513

\ref Vidal-Madjar, A., Ferlet, R., Lemoine, M. and the Fuse team, 1998, in
"Structure and evolution of the intergalactic medium from QSO absorption
lines"
ed. P. Petitjean and S. Charlot, ed. Fronti\`eres, p. 355

\ref von Montigny, C. et al., 1995, ApJ 440, 525

\ref Wampler, E.J. et al., 1996, A\&A 316, 33

\ref Webb, J.K. et al., 1997, Nature 397, 250
}
\end{document}